# Printed high-mobility *p*-type buffer layers on perovskite photovoltaics for efficient semi-transparent devices

*Robert A. Jagt[1], Tahmida N. Huq[1], Sam A. Hill[1], Maung Thway[3], Tianyuan Liu[3], Mari Napari[1,2], Bart Roose[4], Krzysztof Gałkowski[4], Weiwei Li[1], Serena Fen Lin[3], Samuel D. Stranks[4,5], Judith L. MacManus-Driscoll[1], Robert L. Z. Hoye[6,1,\*]*

[1] Department of Materials Science and Metallurgy, University of Cambridge, 27 Charles Babbage Road, Cambridge CB3 0FS, UK

[2] Present address: Zepler Institute for Photonics and Nanoelectronics, University of Southampton, University Road, Southampton SO17 1BJ, UK

[3] Solar Energy Research Institute of Singapore, National University of Singapore, Singapore 117574

[4] Department of Physics, Cavendish Laboratory, University of Cambridge, 19 JJ Thomson Avenue, Cambridge CB3 0HE, UK

[5] Department of Chemical Engineering and Biotechnology, University of Cambridge, Philippa Fawcett Drive, Cambridge CB3 0AS, UK

[6] Present address: Department of Materials, Imperial College London, Exhibition Road, London SW7 2AZ, UK

\*Email: r.hoye@imperial.ac.uk

## Abstract

Perovskite solar cells (PSCs) with transparent electrodes can be integrated with existing solar panels in tandem configurations to increase the power conversion efficiency. A critical layer in semi-transparent PSCs is the inorganic buffer layer, which protects the PSC against damage when the transparent electrode is sputtered on top. The development of *n-i-p* structured semi-transparent PSCs has been hampered by the lack of suitable *p*-type buffer layers. In this work we develop a *p*-type $CuO_x$ buffer layer, which can be grown uniformly over the perovskite device without damaging the perovskite or organic charge transport layers, can be grown using industrially scalable techniques and has high hole mobility ($4.3 \pm 2$ $cm^2$ $V^{-1}$ $s^{-1}$), high transmittance (>95%), and a suitable ionisation potential for hole extraction ($5.3 \pm 0.2$ eV). Semi-transparent PSCs with efficiencies up to 16.7% are achieved using the $CuO_x$ buffer layer. Our work demonstrates a new approach to integrate PSCs into tandem configurations, as well as enable the development of other devices that need high quality *p*-type layers.

## Introduction

Over the past decade, lead-halide perovskites have gained significant attention for optoelectronic applications, such as photovoltaics[1], light-emitting diodes[2], lasers[3], X-ray/photon detectors[4,5] and water splitting[6]. These materials have the general formula ABX$_3$, where A is a monovalent cation (*e.g.*, CH$_3$NH$_3^+$), B a divalent cation (Pb$^{2+}$) and X a halide (I$^-$, Br$^-$, Cl$^-$). By changing the composition on the A, B and X sites, the band gap can be tuned from 1.2 eV to above 3.2 eV[7–9]. The ability to finely tune the band gap over a wide range, together with the ability to achieve long diffusion lengths using low-cost solution-based fabrication methods[10], make lead-halide perovskites promising for applications in tandem photovoltaics, both as the top-cell (*e.g.* over silicon)[11] or in all-perovskite tandem solar cells (using a lead-tin perovskite bottom-cell)[7,12]. The tandem configuration requires the perovskite top-cell to have transparent electrodes on both sides (*i.e.*, a semi-transparent device).

A key component in semi-transparent perovskite devices is an inorganic buffer layer that mechanically protects the perovskite sub-cell from damage when the transparent electrode is deposited on top by sputtering[11,13–16]. Unlike thermal evaporation of metal electrodes, sputtering can damage organic charge transport layers and the lead-halide perovskite film itself due to the high kinetic energy of the atomic species (10–10$^2$ eV for sputtering, compared to tens of meV for evaporation)[17]. A wide range of *n*-type oxides have been developed, which have been integrated between the electron transport layer of *p-i-n* structured perovskite top-cells and the sputtered transparent conducting oxide (TCO). These materials include SnO$_2$[13], zinc tin oxide (ZTO)[11], TiO$_2$, ZnO[18], Al-doped ZnO[19] and indium doped ZnO (IZO)[20]. These wide band gap *n*-type oxides have been shown to protect the perovskite top-cells from sputter damage, transmit visible and near-infrared light and enable efficient electron extraction. As a result, efficient semi-transparent perovskite top-cells with low near-infrared optical losses have been achieved, which have been critical for the rapid development of perovskite-based tandems to a certified power conversion efficiency (PCE) of 28.0%, which already exceeds the record PCE of silicon single-junction devices (26.6%)[1].

However, there are currently no *p*-type oxide buffer layers demonstrated in *n-i-p* structured perovskite top-cells to protect the hole transport and perovskite layer from sputter damage. There are several reasons to integrate *n-i-p* structured perovskite solar cells in a semi-transparent architecture. Firstly, single junction *n-i-p* structured solar cells with evaporated opaque electrodes have consistently outperformed their *p-i-n* counterparts (every entry on the NREL chart for the certified record efficiency perovskite solar cell to date is from a *n-i-p* structured device[1]), yet this has so far not been reflected in the performance of semi-transparent devices[21,22]. Secondly, an *n-i-p* structured top-cell is needed in two-terminal tandems with *n*-type silicon bottom cells, which outperform their *p*-type counterparts. Thirdly, *n-i-p* structured semi-transparent perovskite devices are needed for photoanodes in solar-fuel production, which can be combined with the already-developed *p-i-n* structured perovskite photocathodes for bias-free water splitting[23]. Buffer layers to protect *n-i-p* semi-transparent devices are also needed in multi-junction light-emitting diodes (LEDs; *e.g.*, for white-light devices), transparent complementary metal oxide semiconductor (CMOS) circuits, and for performing spectroscopy on devices under operation.

The difficulty in achieving *p*-type semiconductors with equivalent performance to *n*-type semiconductors comes from the fact that typically in oxides the formation energy of donor defects such as oxygen vacancies is low compared to acceptor defects such as cation vacancies, making them *n*-type. Due to the current lack of *p*-type oxide buffer layers, the majority of researchers utilise MoO$_x$ (a high work function *n*-type oxide) as a buffer layer[12,15,16,22,24,25]. However MoO$_x$ reacts strongly with the lead-halide perovskite, hampering long-term stability[26,27]. Researchers are actively searching for alternatives, but these efforts have mainly focussed on other high work function *n*-type oxides. Raiford *et al.* recently investigated *n*-type VO$_x$ as a buffer layer. They obtained semi-transparent *n-i-p* devices with 1000 h stability, but the maximum PCE was 14%[28]. Another option is *n*-type WO$_x$, but this has not yet been tested as a sputter buffer layer[26]. Hou *et al.* reported that it was necessary to dope WO$_x$ with Ta, which decreased near-infrared transmittance and would therefore result in increased parasitic optical losses to the bottom cell in a tandem.

Owing to the limitations of high work function *n*-type oxides, developing *p*-type oxide buffer layers is critical. Viable potential materials can be found by looking to the *p*-type oxides that have been used as hole transport layers beneath the perovskite in single-junction devices. The most popular is NiO$_x$[29,30]. However, this is typically grown by energetic processes (*e.g.*, reactive sputtering, chemical vapor deposition at temperatures >250 ºC or plasma assisted ALD), which are not compatible with the lead-halide perovskite[31,32]. Another option is Cu$_2$O. Unlike most metal oxides, which have low hole mobilities due to localised O 2$p$ orbitals, the valence band density of states of Cu$_2$O is formed by the hybridisation of Cu 3$d$ and O 2$p$ orbitals[33]. The greater dispersion enables higher hole mobilities, which have exceeded 100 cm$^2$ V$^{-1}$ s$^{-1}$ in single crystals[34]. Cu$_2$O in direct contact with CH$_3$NH$_3$PbI$_3$ as a hole transport layer beneath the perovskite does not accelerate degradation[35], unlike MoO$_x$[26,27]. Several options exist to grow thin films of Cu$_2$O, such as spin coating Cu$_2$O nanoparticles, sputtering[36], thermal evaporation[37], pulsed laser deposition[34], atomic layer deposition (ALD)[38] or atmospheric pressure chemical vapour deposition (AP-CVD)[39–41]. Due to the limited thermal stability of perovskites the deposition method must be able to achieve the film requirements (high mobility, high transmittance, pinhole-free) at low temperatures[28,42].

In this work we report on using *p*-type CuO$_x$ grown at low temperatures as a buffer layer for *n-i-p* structured perovskite stacks for semi-transparent solar cells. We investigate the influence of the growth conditions on the structural, electrical and electronic properties of the films and their suitability for buffer layer applications. In particular, we use structural and spectroscopic characterisation to probe whether CuOx degrades the bulk or interfaces of the perovskite films. Finally, we test the CuO$_x$ layer in semi-transparent devices, and focus on understanding whether sputter damage to the perovskite and organic hole transport layers can be further reduced through thermalized deposition of the TCO, whilst not sacrificing the carrier and optical properties of the TCO. These semi-transparent devices are used as top-cells in four-terminal tandems with two types of silicon bottom cells that are industrially-relevant: *p*-type passivated emitter rear contact (PERC) and *n*-type silicon cells with polycrystalline silicon passivation[43,44].

## Results

The CuO$_x$ layers were deposited by AP-CVD. Allyloxytrimethylsilyl Hexafluoroacetylacetonate Copper(I) (Cupraselect®) was used as the metal precursor and water vapour as the oxidant. The deposition temperatures were varied between 100 °C and 150 °C (refer to Methods). Based on prior work with *n*-type AP-CVD oxides grown over lead-halide perovskites, this is the temperature range that was expected to be compatible with the perovskites[45]. By changing the exposure time per cycle, it was found that there was a linear dependence of the growth rate with exposure time, showing the growth mechanism to be chemical vapour deposition rather than atomic layer deposition. The growth rate at 100 °C and 150 °C were similar, at 1.30 nm min$^{-1}$ and 1.34 nm min$^{-1}$ respectively (Supplementary Fig.1), which is an order of magnitude higher than for conventional ALD[39]. The films can be grown over large areas (>6 cm$^2$ deposition area, Figure 1a), with a thickness variation of only 5% (one standard deviation relative to the mean thickness, Figure 1b).

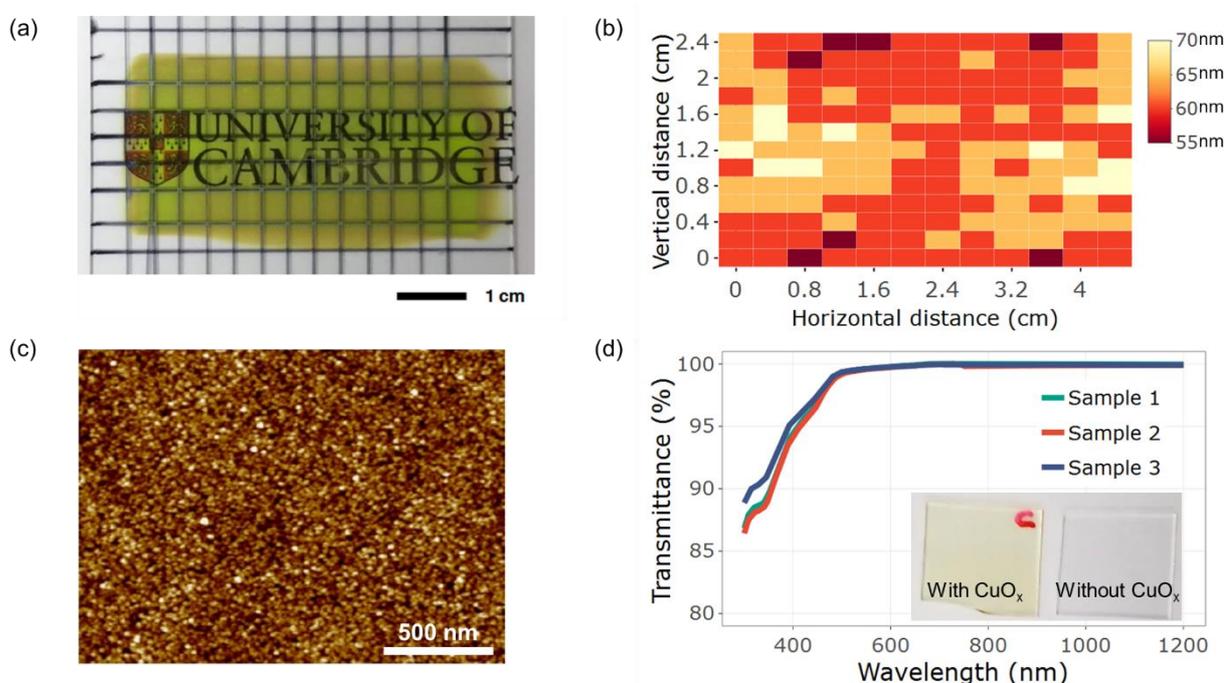

**Figure 1.** Properties of AP-CVD CuO$_x$ grown onto glass. Analysis of 60 nm thick CuO$_x$: (a) Photograph of 60 nm CuO$_x$ grown at 100 °C onto glass, and (b) the measured thicknesses over the deposition area. Analysis of 3 nm (thickness determined from growth rate) thick CuO$_x$: (c) Atomic force microscopy image of CuO$_x$ grown onto glass (root mean square roughness 0.64 nm) and (d) corresponding measured transmittance. The transmittance of the film is corrected for the absorption of the glass. The inset shows photographs of borosilicate glass with (left) and without (right) the CuO$_x$ layer.

The phase stability of Cu$_2$O is narrow and careful control of the processing conditions is required to avoid impurity phases such as Cu metal or CuO[46,47]. X-ray diffraction measurements showed the films (50 nm thick) to be phase-pure cubic Cu$_2$O, with no crystalline CuO phase-impurities detected (Supplementary Fig. 2). X-ray photoemission spectroscopy (XPS) measurements confirm the films to contain Cu in the 1+ oxidation state, with no 2+ satellites observed on the surface (Supplementary

Fig. 3). After sputtering through the surface contaminant layer, we can still detect a very small C 1s core level peak, probably originating from the metal precursor. We measured the bulk composition in detail using time-of-flight elastic recoil detection analysis (ToF-ERDA) and found the C content to be low, at 1.9±0.2 at.% (100 °C growth temperature), decreasing to 0.7±0.1 at.% for films grown at 150 °C (Supplementary Table 1 and Supplementary Fig.4). We measured the bulk ratio of O to Cu (*i.e.*, x value in $CuO_x$) to be 0.4 across all deposition temperatures (specifically 0.39±0.03 for the film deposited at 100 °C), which is sub-stoichiometric for $Cu_2O$ (or $CuO_{0.5}$).

We investigated whether the growth of $CuO_x$ onto perovskite devices results in bulk structural damage. To test this we measured the XRD pattern, before and after $CuO_x$ deposition, of a triple cation lead-halide perovskite film[1] with PTAA[2] on top of a glass substrate. The as-grown triple-cation perovskites have residual lead iodide ($PbI_2$) present, which may be beneficial for passivation purposes[48]. Depositing $CuO_x$ for 100 cycles (approximately 3 nm based on the growth rate, total deposition time <2 min) at 100 °C over the perovskite/PTAA film resulted in no significant increase in the bulk $PbI_2$ peak intensity (Supplementary Fig. 5). However, at 125 °C, the $PbI_2$ peak became significantly more intense, indicating decomposition of the perovskite. We therefore focussed on growth at 100 °C.

To study the morphology of the AP-CVD $CuO_x$ film, we deposited onto $O_2$ plasma treated glass at 100 °C. The atomic force microscopy (AFM) image shows the film to be devoid of pinholes (Fig.1c), with a root mean square roughness of 0.64 nm compared to 0.21 nm for the bare glass (Supplementary Fig. 1). Owing to the thin nature of these films, the transmittance was >95% over the entire visible and near-infrared wavelength range (Fig. 1d), and was close to 100% for near-infrared wavelengths. The $CuO_x$ films grown for <2 min in atmospheric conditions are therefore dense, pinhole-free and give negligible parasitic optical losses.

Hall measurements of 100 nm $CuO_x$ grown on glass at 100 °C confirmed the films to be *p*-type with an average measured Hall mobility of 4.3 ± 2 $cm^2 V^{-1} s^{-1}$ and a carrier concentration of (2.0 ± 1.1)×$10^{15}$ $cm^{-3}$. The Hall mobility is orders of magnitude higher than that of the organic hole transport layers, such as PTAA and Spiro-OMeTAD, used in *n-i-p* perovskite solar cells[49], with no need for doping.

We investigated whether $CuO_x$ grown at 100 °C has a suitable band alignment with perovskite devices. $CuO_x$ was grown onto silicon substrates coated with Au. Using Kelvin probe measurements, we found the work function to be a stable value of 4.9 ± 0.1 eV (Supplemental Fig. 6). Combined with a Fermi level to valence band offset of 0.40 ± 0.1 eV from XPS measurements (Supplemental Fig. 6), the ionisation potential of $CuO_x$ was found to be 5.3 ± 0.2 eV. This is very well matched with the highest occupied molecular orbital (HOMO) of PTAA and the perovskite valence band maximum, making the AP-CVD $CuO_x$ well suited as buffer layer in *n-i-p* structured perovskite devices.

---

[1] ($Cs_{0.05}(MA_{0.17}FA_{0.83})_{0.95}Pb(I_{0.83}Br_{0.17})_3$), MA = $CH_3NH_3^+$, FA = $CH(NH_2)_2^+$.
[2] poly[bis(4-phenyl)(2,4,6-trimethylphenyl)amine]

To determine whether depositing $CuO_x$ on perovskites and perovskite devices resulted in increased non-radiative recombination, we measured the photoluminescence (PL) decay. Notably, directly depositing $CuO_x$ on lead-halide perovskites at 100 °C results in similar PL decay kinetics to the pristine perovskite itself (Fig. 2a), suggesting the deposition of $CuO_x$ does not degrade the surface of the perovskite or introduce traps at the interface. Similarly, the PL decay kinetics of perovskite/PTAA bilayers were unchanged with the $CuO_x$ layer added (Fig. 2b), showing $CuO_x$ to also not damage the thermally-sensitive PTAA layer, in spite of growth in air. Lastly the PL decay was not quenched when the tin-doped indium oxide (ITO) was deposited on top (Fig. 2c & 2d). The absence of quenching of the PL at open circuit voltage imply no new introduction of electric fields or electronic interfaces, both being positive in optoelectronic terms.

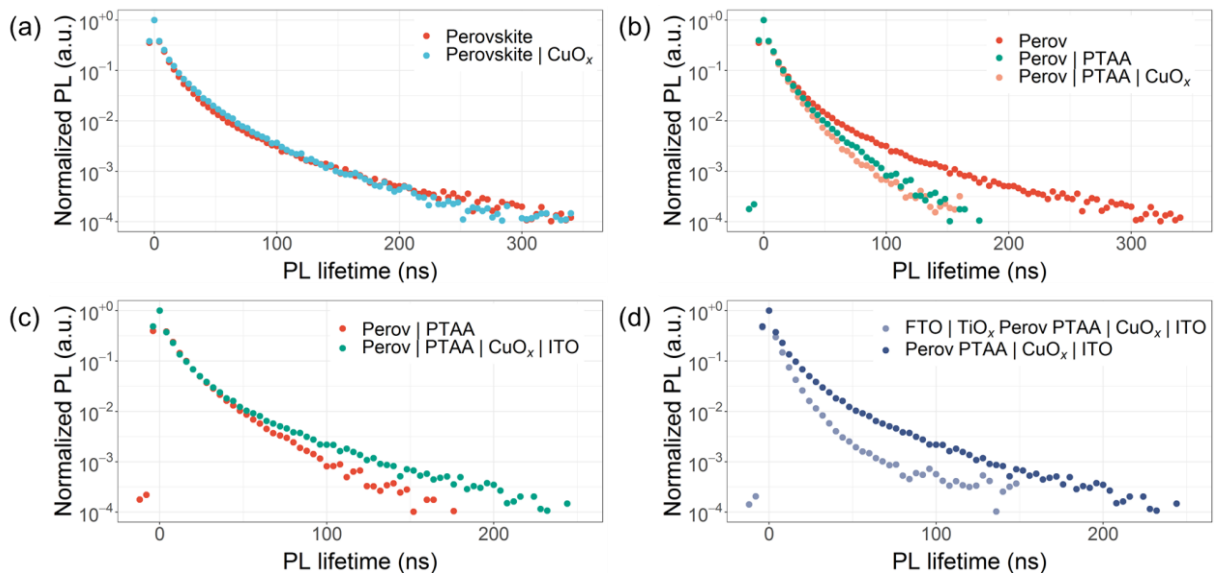

**Figure 2.** Measured photoluminescence lifetime for different device stack configurations (device configuration indicated in the figure legends).

The transparent electrode ITO was deposited by DC magnetron sputtering. For details see the Methods section. The 320 nm thick ITO film had a measured Hall mobility of 35.6 $cm^2\,V^{-1}\,s^{-1}$ and a carrier concentration of $9.0 \times 10^{20}$ $cm^{-3}$ and a sheet resistance of 9.2 Ω/□.

Having found the $CuO_x$ layer to be of high quality and not damage the perovskite/PTAA stack, we investigated its effect on device performance and whether it can be used to protect against sputter damage. A cross-sectional SEM image of the semi-transparent devices with the corresponding band diagram is shown in Fig. 3a & 3b. The top-down scanning electron microscopy images are shown in Supplementary Fig. 8. We observed the morphology of the perovskite/PTAA stack without and with the $CuO_x$ to be the same, suggesting the $CuO_x$ layer to be conformal. This is consistent with previous work, in which it was shown through transmission electron microscopy images that $CuO_x$ can be grown conformal to high-aspect ratio nanorods[40].

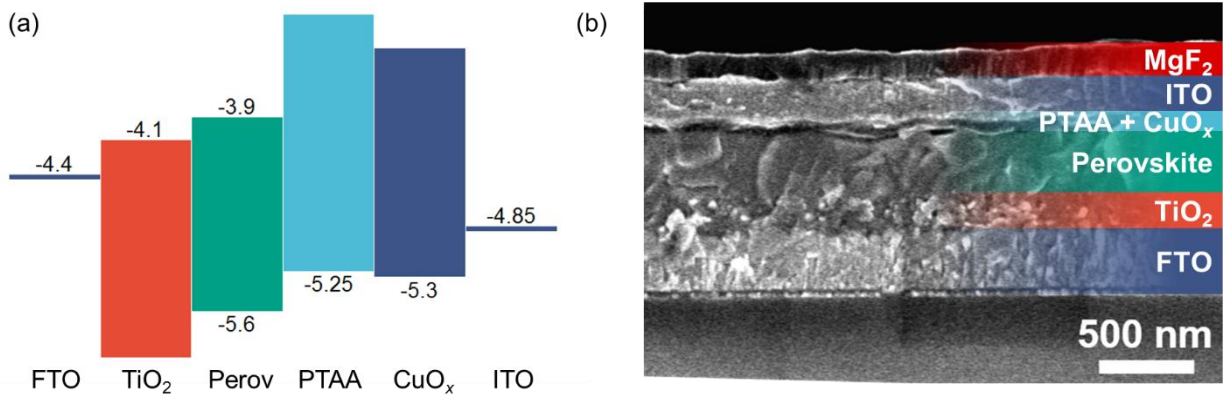

**Figure 3.** (a) Band structure of the semi-transparent device. The CuO$_x$ and ITO band positions were measured by Kelvin probe and X-ray photoemission spectroscopy. The band positions of the other layers were obtained from the literature[49,50]. (b) Cross-sectional scanning electron image of the semi-transparent device. The device structure was MgF$_2$/glass/FTO/c-TiO$_2$/mp-TiO$_2$/Cs$_{0.05}$(MA$_{0.17}$FA$_{0.83}$)$_{0.95}$Pb(I$_{0.83}$Br$_{0.17}$)$_3$/PTAA/CuO$_x$/ITO/MgF$_2$.

Next we investigate the performance of the CuO$_x$ buffer layer in devices. The control *n-i-p* structured devices with opaque Au top electrodes had a standard efficiency of up to 19.9% (median of 17.4%, Supplementary Fig. 8). When the Au top electrode was replaced by ITO directly sputtered on the PTAA layer, the device efficiency dropped significantly. The maximum achieved efficiency without a buffer layer was 12.5%. The low efficiency was mostly due to a low fill factor (< 61%, see Fig. 4a). However, when we added a buffer layer of 3 nm AP-CVD CuO$_x$ in our perovskite devices, we achieved reproducibly rectifying devices, with efficiencies up to 16.7% (Fig. 4, Table 1 and Supplementary Fig. 8). For the champion device the integrated EQE (20.3 mA cm$^{-2}$) matched with the measured short circuit current density (20.6 mA cm$^{-2}$, see Fig. 4d). The performance of our semi-transparent devices is comparable to the best opaque-electrode devices using Cu/Cu$_2$O by Chen *et al.*[51], and is also comparable to the state-of-the-art performance in semi-transparent perovskite devices[11,16,21,22,52–54]. The AP-CVD CuO$_x$ therefore allows protection of the perovskite from sputter damage with thin organic charge transport layers, and high transmittance and performance to all be simultaneously achieved.

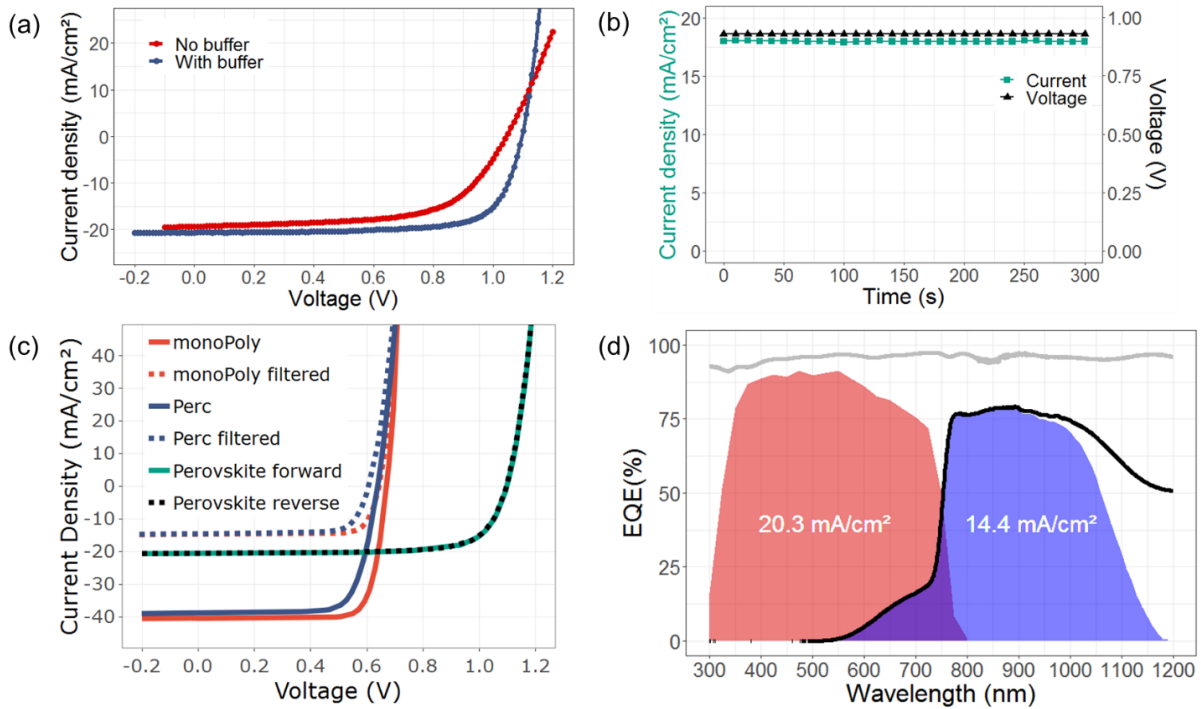

**Figure 4.** (a) Measured reverse current density – voltage curve of the best devices with and without a CuO$_x$ buffer layer. (b) Maximum power point tracking of the current density and voltage of the champion device with a CuO$_x$ buffer layer. (c) Measured current density – voltage curves of the semi-transparent perovskite solar cell and the unfiltered and filtered Si bottom cell. (d) Measured external quantum efficiency (EQE) of the perovskite solar cells (red), transmittance of solar cell device stack (black), EQE of filtered bottom Si cell (blue) and reflection (grey, dotted).

We applied our efficient semi-transparent perovskite devices as top-cells in four-terminal tandems with silicon. The total efficiency of the tandem is determined by the combined efficiency of the top and bottom cell. More efficient bottom cells usually result in lower efficiency gains despite an overall higher tandem efficiency. We fabricated a *p*-type passivated emitter rear content (PERC) bottom cell, which is commonly used in industry, with an efficiency of 18.5% (Table 1). The PERC cell is optimised to function as a bottom cell in a four terminal tandem[43]. The sunlight was filtered using 4 cm$^2$ semi-transparent perovskite devices, and the measured bottom cell efficiency was 6.70%, which is close to our calculated value based on the transmittance of the perovskite filter. Overall, adding our perovskite top-cell results in an efficiency gain of 4.95% in the four-terminal tandem compared to the unfiltered Si cell.

We also made four-terminal tandems with *n*-type silicon bottom cells with passivating poly-silicon layers[44]. The n-type monoPoly$^{TM}$ silicon solar cells are highly industrially relevant due to their simple fabrication process. The structure can be realized with a lean 7 steps processing sequence [44]. Owing to the higher performance of the *n*-type bottom cells, we achieved a four-terminal tandem efficiency of 24.3%, with an overall gain of 2.78% over the silicon bottom cell (Table 1). This shows the future potential of making two-terminal tandems with these *n*-type bottom cells, in which it is critical that the top-cell is *n-i-p* structured. We also note that if we were to use the record-efficiency silicon heterojunction interdigitated back contact solar cell as our bottom cell[55], we would expect to achieve a tandem efficiency of up to 27% (calculation details in Supplementary). This would be comparable to the highest four-terminal tandem efficiency reported for any perovskite top-cell structure[16,56].

**Table 1.** Summary of champion device with measured device data for *p*-type Perc Si and n-type monoPolyTM c-Si solar cells.

|  | $J_{sc}$ (mA cm$^{-2}$) | $V_{oc}$ (mV) | FF(%) | PCE (%) |
|---|---|---|---|---|
| Semi-transparent Perovskite | 20.63 | 1100.0 | 73.66 | 16.72 |
| c-Si PERC | 38.78 | 632.7 | 75.26 | 18.47 |
| c-Si PERC filtered | 14.54 | 597.9 | 77.09 | 6.70 |
| Tandem with c-Si PERC |  |  |  | 23.42 |
| monoPoly$^{TM}$ c-Si | 40.39 | 664.2 | 80.45 | 21.58 |
| monoPoly$^{TM}$ c-Si filtered | 14.73 | 634.1 | 81.84 | 7.64 |
| Tandem with monoPoly$^{TM}$ c-Si |  |  |  | 24.36 |

**Discussion**

Sputter damage can occur to the perovskite device despite the presence of a buffer layer if the sputtering conditions are not carefully optimised[57]. The damage is due to the high kinetic energy of the sputtered particles. The conductance and transmittance of the TCO and the extent of damage to the solar cells are highly dependent on the deposition power, source to substrate distance and working pressure[58]. Typically TCOs are deposited at low pressure to maximise the carrier concentration[11,13,16,19]. However, decreasing the pressure also increases the kinetic energy of the atomic species arriving at the substrate surface, inducing more sputter damage. Using standard kinetic gas theory it is possible to calculate the distance the atomic species need to travel before their kinetic energy is reduced to the kinetic energy of the background gas. See the supplementary information for more details. This distance is called the thermalization distance and is depicted in Fig. 5 as a function of sputter pressure.

In this work the sputter pressure and source to substrate distance were 2 Pa and 5 cm. As can be seen in Figure 5 the thermalization distance for the various atoms is smaller than the source to substrate distance. Hence the atomic species originating from the target have slowed down to the background gas temperature. The TCO still had high carrier concentration and mobility values of $9.0 \times 10^{20}$ cm$^{-3}$ and 35.6 cm$^2$ V$^{-1}$ s$^{-1}$. In other works, the background pressure is typically set between 0.2 and 0.4 Pa. The thermalization distance of the different atoms is then between 20 and 30 cm and therefore sputtering is done in the ballistic regime where the kinetic energy of the atomic species is still significantly higher than the kinetic energy of the background gas. It is hypothesized that the increased background pressure reduces interfacial damage. In this work depositing at lower pressure caused the TCO films to delaminate due to the increased compressive stresses (see Supplementary Fig. 10). Further work is needed to assess the impact of sputter pressure on interfacial damage between the TCO and transport or buffer layers.

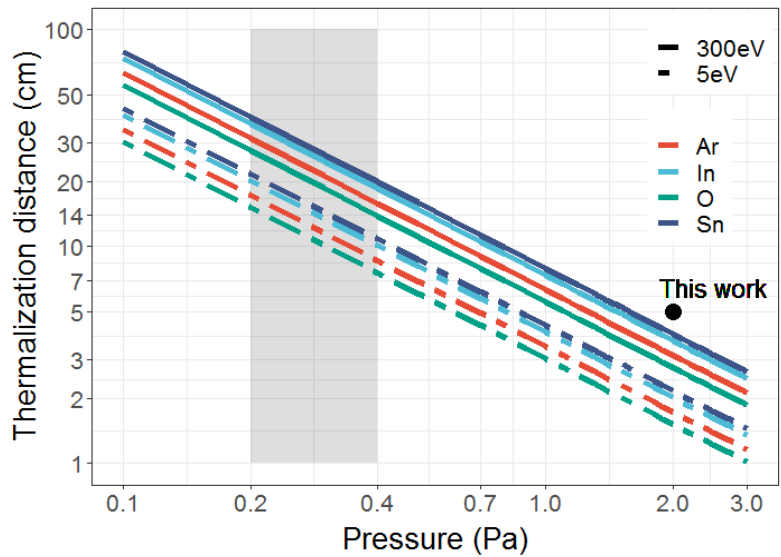

**Figure 5.** Calculated thermalization distance for different atomic species (In, Sn, O, Ar) and different initial kinetic energy (5 eV, 300 eV) as a function of pressure. The grey area indicates typical employed sputtering pressures. The black circle at (2 Pa, 5 cm) indicates the pressure and the source to substrate window at which the ITO is sputtered here.

**Conclusion**

In summary, the current lack of *p*-type buffer layers hampers the development of semi-transparent *n-i-p* structured perovskite devices. In this work we have developed thin high mobility *p*-type $CuO_x$ films, which can be grown at low temperatures. This enabled us to achieve *n-i-p* semi-transparent devices with state-of-the-art performance and high near-infrared transmittance, enabling high efficiency gains when used as top-cells in four-terminal tandems with silicon bottom cells. The *p*-type oxide we used is $CuO_x$ (*x* = 0.39±0.03 from ToF-ERDA measurements) grown by AP-CVD with a high deposition rate (1.3 nm min$^{-1}$), without vacuum, and with high film uniformity (5% variation over >6 cm$^2$ deposition area) and hole mobility (4.3 ± 2 cm$^2$ V$^{-1}$ s$^{-1}$) without requiring dopants. Thin 3 nm films were found to be pinhole-free with >95% transmittance to visible and near-infrared light. XRD and time-resolve photoluminescence showed that growing $CuO_x$ at 100 °C on perovskite devices resulted in no bulk structural damage or extra non-radiative recombination at the interface. The thin $CuO_x$ was found to be sufficient to protect the perovskite device from sputter damage. Our work opens up a new avenue to exploit the most efficient perovskite device architecture (*n-i-p*) in semi-transparent devices and in top-cells for tandem photovoltaics. This opens up their use in photoanodes for bias-free solar fuel generation, multi-junction white-light LEDs and other transparent devices and electronic circuits.

**Acknowledgements**

The authors acknowledge KP Technologies Ltd. for providing the Kelvin probe. Prof. Timo Sajavaara is acknowledged for providing access and support in University of Jyväskylä Accelerator Laboratory. R.A.J. acknowledges funding from an EPSRC Department Training Partnership studentship (No: EP/N509620/1), as well as Bill Welland. T.N.H. acknowledges funding from the EPSRC Centre for Doctoral Training in Graphene Technology (No. EP/L016087/1) and the Aziz Foundation. W.-W.L. and J.L.M.-D. acknowledge support from the EPSRC (Nos.: EP/L011700/1, EP/N004272/10), and the Isaac Newton Trust (Minute 13.38(k)). M.N. and J.L.M.-D. acknowledge financial support from EPSRC (No. EP/P027032/1). S. D. S. acknowledges support from the Royal Society and Tata Group (UF150033). R.L.Z.H. acknowledges support from the Royal Academy of Engineering under the Research Fellowship scheme (No.: RF\201718\1701), the Centre of Advanced Materials for Integrated Energy Systems (EPSRC Grant No. EP/P007767/1), the Isaac Newton Trust (Minute 19.07(d)), and the Kim and Juliana Silverman Research Fellowship at Downing College, Cambridge.

**Author contributions**

R.A.J performed the device fabrication, film growth and characterization with the help of T.N.H. and S.H. M.T. and T.L. did the tandem device measurements. W.L. performed XPS measurements. M.N. did the ToF-ERDA measurement and data analysis of the $CuO_x$ films and assisted in the Hall effect measurements. B.R. assisted in growth of the perovskite devices. K.G, supervised by S.D.S, assisted with the Kelvin Probe measurements. R.L.Z.H. conceived of the project and initiated the development of the growth of $CuO_x$ by AP-CVD. All authors contributed to discussing and writing the paper.

**Declaration of Interests**

The authors declare no competing interests.

**Additional Information**

Supplementary Information accompanies this paper at